\documentclass{frontiersFPHY} % for Physics and Applied Mathematics and Statistics articles

\usepackage{url,hyperref,lineno,microtype}
\usepackage[onehalfspacing]{setspace}
\usepackage{dirtytalk}

\def\keyFont{\fontsize{8}{11}\helveticabold }
\def\firstAuthorLast{Vidgen {et~al.}} %use et al only if is more than 1 author
\def\Authors{Bertie Vidgen\,$^{1}$ and Taha Yasseri\,$^{1,*}$}
\def\Address{$^{1}$Oxford Internet Institute, University of Oxford, Oxford, UK }
\def\corrAuthor{Taha Yasseri}
\def\corrAddress{Oxford Internet Institute, University of Oxford, 1 St Giles, Oxford, OX13JS, UK}
\def\corrEmail{taha.yasseri@oii.ox.ac.uk}

\begin{document}
\onecolumn
\firstpage{1}

\title[P-values: misunderstood and misused]{P-values: misunderstood and misused} 

\author[\firstAuthorLast ]{\Authors} %This field will be automatically populated
\address{} %This field will be automatically populated
\correspondance{} %This field will be automatically populated

\extraAuth{}% If there are more than 1 corresponding author, comment this line and uncomment the next one.
%\extraAuth{corresponding Author2 \\ Laboratory X2, Institute X2, Department X2, Organization X2, Street X2, City X2 , State XX2 (only USA, Canada and Australia), Zip Code2, X2 Country X2, email2@uni2.edu}

\maketitle

%%%%%%%%%%%%%%%%%%%%%%%%%%%%%%%%%%%%%%%%%%%%%%%%%%%%%%%%%%%%%%%%%%%%%%%%%%%%%%%%%%%%%%%%%%%%%%%%%%%%%%%%%%%%%%%%%%%%%%%%%%%%%%%%%%%%%%%%%%%%%%%%%%%%%%%%%%%%%%%%%%%%%%%%%%%%%%%%%%%%%%%%%%%%%%%%%%%%%%%%%%%%%%%%%%%%%%%%%%%%%%%%%%%%%%%
%%% The sections below are for reference only.
%%%
%%% For Original Research Articles, Clinical Trial Articles, and Technology Reports the section headings should be those appropriate for your field and the research itself. It is recommended to organize your manuscript in the
%%% following sections or their equivalents for your field:
%%% Abstract, Introduction, Material and Methods, Results, and Discussion.
%%% Please note that the Material and Methods section can be placed in any of the following ways: before Results, before Discussion or after Discussion.
%%%
%%%For information about Clinical Trial Registration, please go to http://www.frontiersin.org/about/AuthorGuidelines#ClinicalTrialRegistration
%%%
%%% For Clinical Case Studies the following sections are mandatory: Abstract, Introduction, Background, Discussion, and Concluding Remarks.
%%%
%%% For all other article types there are no mandatory sections.
%%%%%%%%%%%%%%%%%%%%%%%%%%%%%%%%%%%%%%%%%%%%%%%%%%%%%%%%%%%%%%%%%%%%%%%%%%%%%%%%%%%%%%%%%%%%%%%%%%%%%%%%%%%%%%%%%%%%%%%%%%%%%%%%%%%%%%%%%%%%%%%%%%%%%%%%%%%%%%%%%%%%%%%%%%%%%%%%%%%%%%%%%%%%%%%%%%%%%%%%%%%%%%%%%%%%%%%%%%%%%%%%%%%%%%%

\begin{abstract}

%%% Leave the Abstract empty if your article falls under any of the following categories: Editorial Book Review, Commentary, Field Grand Challenge, Opinion or specialty Grand Challenge.
\section{}
%As a primary goal, the abstract should render the general significance and conceptual advance of the work clearly accessible to a broad readership. References should not be cited in the abstract.
P-values are widely used in both the social and natural sciences to quantify the statistical significance of observed results. The recent surge of big data research has made the p-value an even more popular tool to test the significance of a study. However, substantial literature has been produced critiquing how p-values are used and understood. In this paper we review this recent critical literature, much of which is routed in the life sciences, and consider its implications for social scientific research. We provide a coherent picture of what the main criticisms are, and draw together and disambiguate common themes. In particular, we explain how the False Discovery Rate is calculated, and how this differs from a p-value. We also make explicit the Bayesian nature of many recent criticisms, a dimension that is often underplayed or ignored. We conclude by identifying practical steps to help remediate some of the concerns identified. We recommend that (i) far lower significance levels are used, such as $0.01$ or $0.001$, and (ii) p-values are interpreted contextually, and situated within both the findings of the individual study and the broader field of inquiry (through, for example, meta-analyses). 

\tiny
 \keyFont{ \section{Keywords:} p-value, statistics, significance, p-hacking, prevalence, Bayes, big data } %All article types: you may provide up to 8 keywords; at least 5 are mandatory.
\end{abstract}

\section{Introduction}

% For Original Research Articles, Clinical Trial Articles, and Technology Reports the introduction should be succinct, with no subheadings.
%
% For Clinical Case Studies the Introduction should include symptoms at presentation, physical exams and lab results.
%
P-values are widely used in both the social and natural sciences to quantify the statistical significance of observed results. Obtaining a p-value that indicates \say{statistical significance} is often a requirement for publishing in a top journal. The emergence of computational social science, which relies mostly on analyzing large scale datasets, has increased the popularity of p-values even further. However, critics contend that p-values are routinely misunderstood and misused by many practitioners, and that even when understood correctly they are an ineffective metric: the standard significance level of $0.05$ produces an overall False Discovery Rate that is far higher, more like 30\%. Others argue that p-values can be easily \say{hacked} to indicate statistical significance when none exists, and that they encourage the selective reporting of only positive results.

Considerable research exists into how p-values are (mis)used, 
\citep[e.g.][]{ioannidis2005,ziliak2008}. In this paper we review the recent critical literature on p-values, much of which is routed in the life sciences, and consider its implications for social scientific research. We provide a coherent picture of what the main criticisms are, and draw together and disambiguate common themes. In particular, we explain how the False Discovery Rate is calculated, and how this differs from a p-value. We also make explicit the Bayesian nature of many recent criticisms. In the final section we identify practical steps to help remediate some of the concerns identified.

P-values are used in Null Hypothesis Significance Testing (NHST) to decide whether to accept or reject a null hypothesis (which typically states that there is no underlying relationship between two variables). If the null hypothesis is rejected, this gives grounds for accepting the alternative hypothesis (that a relationship does exist between two variables). The p-value quantifies the probability of observing results at least as extreme as the ones observed given that the null hypothesis is true. It is then compared against a pre-determined significance level ($\alpha$). If the reported p-value is smaller than $\alpha$ the result is considered statistically significant. Typically, in the social sciences $\alpha$ is set at $0.05$. Other commonly used significance levels are $0.01$ and $0.001$. 

In his seminal paper, \say{The Earth is Round $(p<.05)$} Cohen argues that NHST is highly flawed: it is relatively easy to achieve results that can be labelled significant when a \say{nil} hypothesis (where the effect size of $H_0$ is set at zero) is used rather than a true \say{null} hypothesis (where the direction of the effect, or even the effect size, is specified) \cite{cohen1994}. This problem is particularly acute in the context of \say{big data} exploratory studies, where researchers only seek statistical associations rather than causal relationships. If a large enough number of variables are examined, effectively meaning that a large number of null/alternative hypotheses are specified, then it is highly likely that at least some 'statistically significant' results will be identified, irrespective of whether the underlying relationships are truly meaningful. As big data approaches become more common this issue will become both far more pertinent and problematic, with the robustness of many \say{statistically significant} findings being highly limited. 

Lew argues that the central problem with NHST is reflected in its hybrid name, which is a combination of (i) hypothesis testing and (ii) significance testing  \cite{lew2013}. In significance testing, first developed by Ronald Fisher in the 1920s, the p-value provides an index of the evidence against the null hypothesis. Originally, Fisher only intended for the p-value to establish whether further research into a phenomenon could be justified. He saw it as one bit of evidence to either support or challenge accepting the null hypothesis, rather than as conclusive evidence of significance \cite{fisher1925}; see also \cite{sterne2001,nuzzo2014}. In contrast, hypothesis tests, developed separately by Neyman and Pearson, replace Fisher's subjectivist interpretation of the p-value with a hard and fast \say{decision rule}: when the p-value is less than $\alpha$, the null can be rejected and {\bf the alternative hypothesis accepted}. Though this approach is simpler to apply and understand, a crucial stipulation of it is that a precise alternative hypothesis must be specified \cite{sterne2001}. This means indicating what the expected effect size is (thereby setting a nil rather than a null hypothesis) \textemdash something that most researchers rarely do \cite{cohen1994}. %Furthermore, it is worth noting that Neyman and Pearson \cite{neyman1933} arguably did not envisage that hypothesis tests would be used as the sole determinants of statistical significance, writing \say{as far as a particular hypothesis is concerned, no test based upon the theory of probability can by itself provide any valuable evidence of the truth or falsehood of that hypothesis.} \citep[p. 290--291]{neyman1933}. 

Though hypothesis tests and significance tests are distinct statistical procedures, and there is much disagreement about whether they can be reconciled into one coherent framework, NHST is widely used as a pragmatic amalgam for conducting research \cite{berger2003,hurlbert2009}. Hulbert and Lombardi argue that one of the biggest issues with NHST is that it encourages the use of terminology such as {\it significant/nonsignificant}. This dichotomizes the p-value on an arbitrary basis, and converts a probability into a certainty. This is unhelpful when the purpose of using statistics, as is typically the case in academic studies, is to weigh up evidence incrementally rather than make an immediate decision \citep[p. 315]{hurlbert2009}. Hulbert and Lombardi's analysis suggests that the real problem lies not with p-values, but with $\alpha$ and how this has led to p-values being interpreted dichotomously: too much importance is attached to the arbitrary cutoff $\alpha \leq 0.05$. 

\section{The False Discovery Rate} \label{sec:FDR}

A p-value of $0.05$ is normally interpreted to mean that there is a 1 in 20 chance that the observed results are nonsignificant, having occurred even though no underlying relationship exists. Most people then think that the overall proportion of results that are false positives is also $0.05$. However, this interpretation confuses the p-value (which, in the long run, will approximately correspond to the {\it type I error rate}) with the False Discovery Rate (FDR). The FDR is what people usually mean when they refer to the error rate: it is the proportion of reported discoveries that are false positives. Though $0.05$ might seem a reasonable level of inaccuracy, a type I error rate of $0.05$ will likely produce an FDR that is far higher, easily 30\% or more. The formula for FDR is:

\begin{equation}
 \frac{\rm False~ Positives}{\rm True~ Positives + False~ Positives}.
\end{equation}

Calculating the number of true positives and false positives requires knowing more than just the type I error rate, but also (i) the statistical power, or \say{sensitivity}, of tests and (ii) the prevalence of effects \cite{colquhoun2014}. Statistical power is the probability that each test will correctly reject the null hypothesis when the alternative hypothesis is true. As such, tests with higher power are more likely to correctly record real effects. Prevalence is the number of effects, out of all the effects that are tested for, that actually exist in the real world. In the FDR calculation it determines the weighting given to the power and the type I error rate. Low prevalence contributes to a higher FDR as it increases the likelihood that false positives will be recorded. The calculation for FDR therefore is: 
\begin{equation}
 %\frac{ N \rm \times (1-Prevalence) \times Type~I~error~rate}{N {\rm \times Prevalence \times Power} + N \rm \times (1-Prevalence) \times Type~I~error~rate,
 \frac{ \rm (1-Prevalence) \times Type~I~error~rate}{\rm Prevalence \times Power +  (1-Prevalence) \times Type~I~error~rate}.
\end{equation}

The percentage of reported positives that are actually true is called the Positive Predictive Value (PPV). The PPV and FDR are inversely related, such that a higher PPV necessarily means a lower FDR. To calculate the FDR we subtract the PPV from 1. If there are no false positives then $\rm PPV = 1$ and ${\rm FDR} = 0$. \textbf{Table \ref{tab:pvalue}} shows how low prevalence of effects, low power, and a high type I error rate all contribute to a high FDR.

\begin{table}[!t]
\textbf{\refstepcounter{table}\label{tab:pvalue} Table \arabic{table}. }{Greater prevalence, greater power, and a lower Type I error rate reduce the FDR}

\processtable{ }
{\begin{tabular}{lllll}\toprule
 Prevalence & Power & Type I error rate & FDR \\\midrule
0.01&	0.8&	0.05&	0.86\\
0.1&	0.8&	0.05&	0.36\\
0.5&	0.8&	0.05&	0.06\\\midrule
0.1&	0.2&	0.05&	0.69\\
0.1&	0.5&	0.05&	0.47\\\midrule
0.1&	0.8&	0.01&	0.10\\
0.1&	0.8&	0.001&	0.01\\\botrule
\end{tabular}}{}
\end{table}

Most estimates of the FDR are surprisingly large; e.g., 50\% \cite{ioannidis2005,biau2010,freedman2015} or 36\% \cite{colquhoun2014}. Jager and Leek more optimistically suggest that it is just 14\% \cite{jager2014}. This lower estimate can be explained somewhat by the fact that they only use p-values reported in abstracts, and have a different algorithm to the other studies. Importantly, they highlight that whilst $\alpha$ is normally set to $0.05$, many studies \textemdash particularly in the life sciences \textemdash achieve p-values far lower than this, meaning that the average type I error rate is less than $\alpha$ of $0.05$ \citep[p. 7]{jager2014}. Counterbalancing this, however, is Colquhoun's argument that because most studies are not \say{properly designed} (in the sense that treatments are not randomly allocated to groups and in RCTs assessments are not blinded) statistical power will often be far lower than reported \textemdash thereby driving the FDR back up again \cite{colquhoun2014}.

Thus, though difficult to calculate precisely, the evidence suggests that the FDR of findings overall is far higher than $\alpha$ of $0.05$. This suggests that too much trust is placed in current research, much of which is wrong far more often than we think. It is also worth noting that this analysis assumes that researchers do not intentionally misreport or manipulate results to erroneously achieve statistical significance. These phenomena, known as \say{selective reporting} and \say{p-hacking}, are considered separately in \textbf{Section \ref{sec:pb}}.

\section{Prevalence and Bayes}
As noted above, the prevalence of effects significantly impacts the FDR, whereby lower prevalence increases the likelihood that reported effects are false positives. Yet prevalence is not controlled by the researcher and, furthermore, cannot be calculated with any reliable accuracy. There is no way of knowing objectively what the underlying prevalence of real effects is. Indeed, the tools by which we might hope to find out this information (such as NHST) are precisely what have been criticised in the literature surveyed here. Instead, to calculate the FDR, prevalence has to be estimated\footnote{In much of the recent literature it is assumed that prevalence is very low, around 0.1 or 0.2 \cite{ioannidis2005,biau2010,colquhoun2014,freedman2015}}. In this regard, FDR calculations are inherently Bayesian as they require the researcher to quantify their subjective belief about a phenomenon (in this instance, the underlying prevalence of real effects). 

Bayesian theory is an alternative paradigm of statistical inference to frequentism, of which NHST is part of. Whereas frequentists quantify the probability of the data given the null hypothesis ($\rm P(D|H_0)$), Bayesians calculate the probability of the hypothesis given the data ($\rm P(H_1|D)$). Though frequentism is far more widely practiced than Bayesianism, Bayesian inference is more intuitive: it assigns a probability to a hypothesis based on how likely we think it to be true.

The FDR calculations outlined above in \textbf{Section \ref{sec:FDR}} follow a Bayesian logic. First, a probability is assigned to the prior likelihood of a result being false ($1- {\rm prevalence}$). Then, new information (the statistical power and type I error rate) is incorporated to calculate a posterior probability (the FDR). A common criticism against Bayesian methods such as this is that they are insufficiently objective as the prior probability is only a guess. Whilst this is correct, the large number of \say{findings} produced each year, as well as the low rates of replicability \cite{lindsay2015}, suggest that the prevalence of effects is, overall, fairly low.  Another criticism against Bayesian inference is that it is overly conservative: assigning a low value to the prior probability makes it more likely that the posterior probability will also be low \cite{gelman2008}. These criticisms notwithstanding, Bayesian theory offers a useful way of quantifying how likely it is that research findings are true. 

Not all of the authors in the literature reviewed here explicitly state that their arguments are Bayesian. The reason for this is best articulated by Colquhoun, who writes that \say{the description \say{Bayesian} is not wrong but it is not necessary} \citep[p. 5]{colquhoun2014}. The lack of attention paid to Bayes  in Ioannidis' well-regarded early article on p-values is particularly surprising given his use of Bayesian terminology: \say{the probability that a research finding is true depends on the prior probability of it being true (before doing the study)} \citep[p. 696]{ioannidis2005}. This perhaps reflects the uncertain position that Bayesianism holds in most universities, and the acrimonious nature of its relationship with frequentism \cite{mcgrayne2011}. Without commenting on the broader applicability of Bayesian statistical inference, we argue that a Bayesian methodology has great utility in assessing the overall credibility of academic research, and that it has received insufficient attention in previous studies. Here, we have sought to make visible, and to rectify, this oversight. 

\section{Publication Bias: Selective Reporting and P-Hacking}\label{sec:pb}

\begin{figure}[!t]
\begin{center}
\includegraphics[width=8.0cm]{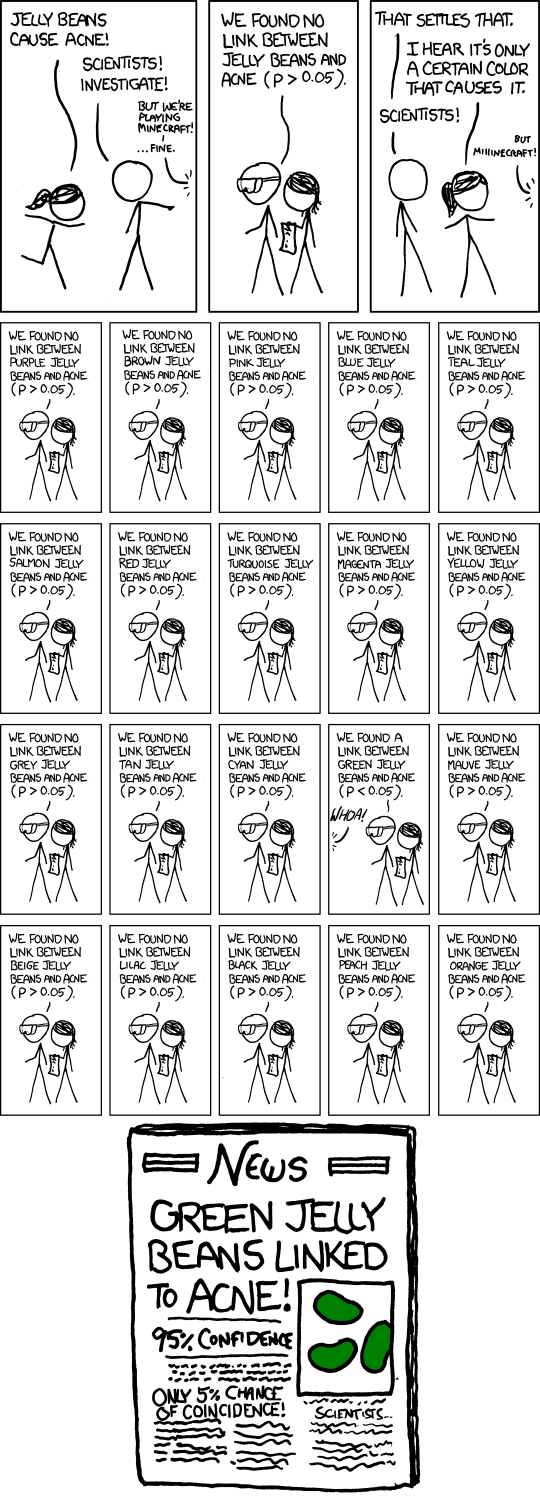}% This is a *.jpg file
\end{center}
 \textbf{\refstepcounter{figure}\label{fig:01} Figure \arabic{figure}.}{\say{Significant}: an illustration of selective reporting and statistical significance from XKCD. Available at http://xkcd.com/882/. Last accessed on 16 February 2016.}
\end{figure}

Selective reporting and p-hacking are two types of researcher-driven publication bias. Selective reporting is where non-significant (but methodologically robust) results are not reported, often because top journals consider them to be less interesting or important \cite{franco2014}. This skews the distribution of reported results towards positive findings, and arguably further increases the pressure on researchers to achieve statistical significance. Another form of publication bias, which also skews results towards positive findings, is called p-hacking. Head et al. define p-hacking as \say{when researchers collect or select data or statistical analyses until nonsignificant results become significant.} \cite{head2015}. This is direct manipulation of results so that, whilst they may not be technically false, they are unrepresentative of the underlying phenomena. See \textbf{Figure \ref{fig:01}} for a satirical illustration. 

Head et al. outline specific mechanisms by which p-values are intentionally \say{hacked}. These include: (i) conducting analyse midway through experiments, (ii) recording many response variables and only deciding which to report postanalysis, (iii) excluding, combining, or splitting treatment groups postanalysis, (iv) including or excluding covariates postanalysis, (vi) stopping data exploration if analysis yields a significant p-value. An excellent demonstration of how p-values can be hacked by manipulating the parameters of an experiment is Christie Aschwanden's interactive \say{Hack Your Way to Scientific Glory} \cite{aschwanden2015}. This simulator, which analyses whether Republicans or Democrats being in office affects the US economy, shows how tests can be manipulated to produce statistically significant results supporting either parties.

In separate papers, Head et al. \cite{head2015}, and  de Winter and Dodou \citep{de2015} each examine the distributions of p-values that are reported in scientific publications in different disciplines. It is reported that there are considerably more studies reporting alpha just below the $0.05$ significance level than above it (and considerably more than would be expected given the number of p-values that occur in other ranges), which suggests that p-hacking is taking place. This core finding is supported by Jager and Leek's study on \say{significant} publications as well \cite{jager2014}. 

\section{What to do}

We argued above that a Bayesian approach is useful to estimate the FDR and assess the overall trustworthiness of academic findings. However, this does not mean that we also hold that Bayesian statistics should replace frequentist statistics more generally in empirical research \citep[see:][]{Simonsohn2014}. In this concluding section we recommend some pragmatic changes to current (frequentist) research practices that could lower the FDR and thus improve the credibility of findings.

Unfortunately, researchers cannot control how prevalent effects are. They only have direct influence over their study's $\alpha$ and its statistical power. Thus, one step to reduce the FDR is to make the norms for these more rigorous, such as by increasing the statistical power of studies. We strongly recommend that $\alpha$ of 0.05 is dropped as a convention, and replaced with a far lower $\alpha$ as standard, such as $0.01$ or $0.001$; see \textbf{Table \ref{tab:pvalue}}. Other suggestions for improving the quality of statistical significance reporting include using confidence intervals \cite[p. 152]{nuzzo2014}. Some have also  called for researchers to focus more on effect sizes than statistical significance \cite{coe2002,sullivan2012}, arguing that statistically significant studies that have negligible effect sizes should be treated with greater scepticism. This is of particular importance in the context of big data studies, where many \say{statistically significant} studies report small effect sizes as the association between the dependent and independent variables is very weak.

Perhaps more important than any specific technical change in how data is analysed is the growing consensus that research processes need to be implemented (and recorded) more transparently. Nuzzo, for example, argues that \say{one of the strongest protections for scientists is to admit everything} \cite[p. 152]{nuzzo2014}. Head et al. also suggest that labelling research as either exploratory or confirmatory will help readers to interpret the results more faithfully \cite[p. 12]{head2015}. Weissgerber et al. encourage researchers to provide \say{a more complete presentation of data}, beyond summary statistics \cite{weissgerber2015}. Improving transparency is particularly important in \say{big} data-mining studies, given that the boundary between data exploration (a legitimate exercise) and p-hacking is often hard to identify, creating significant potential for intentional or unintentional manipulation of results.
Several commentators have recommended that researchers pre-register all studies with initiatives such as the Open Science Framework \cite{head2015,ioannidis2005,nuzzo2014,lindsay2015,peplow2014}. Pre-registering ensures that a record is kept of the proposed method, effect size measurement, and what sort of results will be considered noteworthy. Any deviation from what is initially registered would then need to be justified, which would give the results greater credibility. Journals could also proactively assist researchers to improve transparency by providing platforms on which data and code can be shared, thus allowing external researchers to reproduce a study's findings and trace the method used \cite{head2015}. This would provide academics with the practical means to corroborate or challenge previous findings. 

Scientific knowledge advances through corroboration and incremental progress. In keeping with Fisher's initial view that p-values should be one part of the evidence used when deciding whether to reject the null hypothesis, our final suggestion is that the findings of any single study should always be contextualised within the broader field of research. Thus, we endorse the view offered in a recent editorial of {\it Psychological Science} that we should be extra sceptical about studies where (a) the statistical power is low, (b) the p-value is only slightly below $0.05$, and (c) the result is surprising \cite{lindsay2015}. Normally, findings are only accepted once they have been corroborated through multiple studies, and even in individual studies it is common to \say{triangulate} a result with multiple methods and/or data sets. This offers one way of remediating the problem that even \say{statistically significant} results can be false; if multiple studies find an effect then it is more likely that it truly exists. We therefore also support the collation and organisation of research findings in meta-analyses as these enable researchers to quickly evaluate a large range of relevant evidence.

\section*{Disclosure/Conflict-of-Interest Statement}
%Frontiers follows the recommendations by the International Committee of Medical Journal Editors (http://www.icmje.org/ethical_4conflicts.html) which require that all financial, commercial or other relationships that might be perceived by the academic community as representing a potential conflict of interest must be disclosed. If no such relationship exists, authors will be asked to declare that the research was conducted in the absence of any commercial or financial relationships that could be construed as a potential conflict of interest. When disclosing the potential conflict of interest, the authors need to address the following points:
%•	Did you or your institution at any time receive payment or services from a third party for any aspect of the submitted work?
%•	Please declare financial relationships with entities that could be perceived to influence, or that give the appearance of potentially influencing, what you wrote in the submitted work.
%•	Please declare patents and copyrights, whether pending, issued, licensed and/or receiving royalties relevant to the work.
%•	Please state other relationships or activities that readers could perceive to have influenced, or that give the appearance of potentially influencing, what you wrote in the submitted work.
The authors declare that the research was conducted in the absence of any commercial or financial relationships that could be construed as a potential conflict of interest.

\section*{Author Contributions}
%When determining authorship the following criteria should be observed:
%•	Substantial contributions to the conception or design of the work; or the acquisition, analysis, or interpretation of data for the work; AND
%•	Drafting the work or revising it critically for important intellectual content; AND
%•	Final approval of the version to be published ; AND
%•	Agreement to be accountable for all aspects of the work in ensuring that questions related to the accuracy or integrity of any part of the work are appropriately investigated and resolved.
%Contributors who meet fewer than all 4 of the above criteria for authorship should not be listed as authors, but they should be acknowledged. (http://www.icmje.org/roles_a.html)
All authors listed, have made substantial, direct and intellectual contribution to the work, and approved it for publication.

\section*{Acknowledgments}

For providing useful feedback on the original manuscript we thank Jonathan Bright, Sandra Wachter, Patricia L. Mabry, and Richard Vidgen.

%\section*{Supplemental Data}

%\bibliographystyle{frontiersinSCNS_ENG_HUMS} % for Science, Engineering and Humanities and Social Sciences articles, for Humanities and Social Sciences articles please include page numbers in the in-text citations
\bibliographystyle{frontiersinHLTH&FPHY} % for Health and Physics articles
%\bibliography{pvalue}

%%% Upload the *bib file along with the *tex file and PDF on submission if the bibliography is not in the main *tex file

%%% Use this if adding the figures directly in the mansucript, if so, please remember to also upload the files when submitting your article
%%% There is no need for adding the file termination, as long as you indicate where the file is saved. In the examples below the files (logo1.jpg and logo2.eps) are in the Frontiers LaTeX folder
%%% If using *.tif files convert them to .jpg or .png

%\begin{figure}
%\begin{center}
%\includegraphics[width=10cm]{logo2}% This is an *.eps file
%\end{center}
%\textbf{\refstepcounter{figure}\label{fig:02} Figure \arabic{figure}.}{ Enter the caption for your figure here.  Repeat as  necessary for each of your figures }
%\end{figure}

%%% If you don't add the figures in the LaTeX files, please upload them when submitting the article.

%%% Frontiers will add the figures at the end of the provisional pdf automatically %%%

%%% The use of LaTeX coding to draw Diagrams/Figures/Structures should be avoided. They should be external callouts including graphics.

\end{document}